\documentclass[11pt]{article}
\usepackage{times,aaspp4,astrobib,epsf}

\input{epsf}

\let\oldfootsep=\footnotesep
\def\lsim{\hbox{ \rlap{\raise 0.425ex\hbox{$<$}}\lower 0.65ex\hbox{$\sim$} }}
\def\gsim{\hbox{ \rlap{\raise 0.425ex\hbox{$>$}}\lower 0.65ex\hbox{$\sim$} }}

\def\msun { \rm {M_\odot}} 
\def\rsun { \rm {R_\odot}}

\def\Amax{A_{\rm max}} 
\def\umin{u_{\rm min}}

\def\umin{u_{\rm min}}
\def\t0{t_{\rm 0}}

\def\kpc {\, {\rm kpc}}

\def\dpri{^{\, \prime \prime }}
\def\spose#1{\hbox to 0pt{#1\hss}}
\def\simlt{\mathrel{\spose{\lower 3pt\hbox{$\mathchar"218$}}
     \raise 2.0pt\hbox{$\mathchar"13C$}}}
\def\simgt{\mathrel{\spose{\lower 3pt\hbox{$\mathchar"218$}}
     \raise 2.0pt\hbox{$\mathchar"13E$}}}

\begin{document}

\title{On Planetary Companions to the MACHO-98-BLG-35 Microlens Star}
\author{
  S.H.~Rhie\altaffilmark{1},             
  D.P.~Bennett\altaffilmark{1,2},      
  A.C.~Becker\altaffilmark{2,3},         
  B.A.~Peterson\altaffilmark{4},         
  P.C.~Fragile\altaffilmark{1},           
  B.R.~Johnson\altaffilmark{5},           
  J.L.~Quinn\altaffilmark{1},          
  A.~Crouch\altaffilmark{6},
  J.~Gray\altaffilmark{6},
  L.~King\altaffilmark{1},           
  B.~Messenger\altaffilmark{6},
  S.~Thomson\altaffilmark{6},
\begin{center}
{\bf (The Microlensing Planet Search Collaboration) }\\
\end{center}
}

\author{
I.A.~Bond\altaffilmark{7,8},
F.~Abe\altaffilmark{9},
B.S.~Carter\altaffilmark{10},
R.J.~Dodd\altaffilmark{7,10,16},
J.B.~Hearnshaw\altaffilmark{8},
M.~Honda\altaffilmark{11},
J.~Jugaku\altaffilmark{12},
S.~Kabe\altaffilmark{13},
P.M.~Kilmartin\altaffilmark{7,8},
B.S.~Koribalski\altaffilmark{14},
K.~Masuda\altaffilmark{9},
Y.~Matsubara\altaffilmark{9},
Y.~Muraki\altaffilmark{9},
T.~Nakamura\altaffilmark{15},
G.R.~Nankivell\altaffilmark{10},
S.~Noda\altaffilmark{9},
N.J.~Rattenbury\altaffilmark{7},
M.~Reid\altaffilmark{16,10},
N.J.~Rumsey\altaffilmark{10},
To.~Saito\altaffilmark{17},
H.~Sato\altaffilmark{15},
S.~Sato\altaffilmark{18},
M.~Sekiguchi\altaffilmark{11},
D.J.~Sullivan\altaffilmark{16},
T.~Sumi\altaffilmark{9},
Y.~Watase\altaffilmark{13},
T.~Yanagisawa\altaffilmark{9},
P.C.M.~Yock\altaffilmark{7} and
M.~Yoshizawa\altaffilmark{19}.
}
\begin{center}
{\bf (The MOA Collaboration)}\\
\end{center}

\altaffiltext{1}{Department of Physics, University of Notre Dame, Notre Dame, IN 46556 }
 
\altaffiltext{2}{Center for Particle Astrophysics,
  University of California, Berkeley, CA 94720 }

\altaffiltext{3}{Departments of Astronomy and Physics,
  University of Washington, Seattle, WA 98195 }

\altaffiltext{4}{Mt.~Stromlo and Siding Spring Observatories,
  Australian National University, Weston, ACT 2611, Australia }

\altaffiltext{5}{Tate Laboratory of Physics, University of Minnesota, 
  Minneapolis, MN 55455}

\altaffiltext{6}{Department of Mathematics and Statistics, Monash University,
   Clayton, Victoria 3168, Australia}
  
\altaffiltext{7}{Faculty of Science, University of Auckland, Auckland, New 
   Zealand.}
  
\altaffiltext{8}{Department of Physics and Astronomy, University of 
  Canterbury, Christchurch, New Zealand}
  
\altaffiltext{9}{STE Laboratory, Nagoya University, Nagoya 464, Japan}
  
\altaffiltext{10}{Carter National Observatory, PO Box 2909, Wellington, New Zealand.}
  
\altaffiltext{11}{Institute for Cosmic Ray Research, University of Tokyo, 
   Tokyo 188, Japan.}
  
\altaffiltext{12}{Research Institute of Civilization, 2-29-3 Sakuragaoka, 
Tama-shi 206, Japan.}
  
\altaffiltext{13}{KEK Laboratory, Tsukuba 305, Japan.}
  
\altaffiltext{14}{Australia Telescope National Facility, CSIRO, PO Box 76, 
Epping, NSW 2121, Australia.}
  
\altaffiltext{15}{Department of Physics, Kyoto University, Kyoto 606, Japan.}
  
\altaffiltext{16}{School of Chemical \& Physical Sciences, 
      Victoria University, Wellington, New Zealand.}
  
\altaffiltext{17}{Tokyo Metropolitan College of Aeronautics, 8-52-1 
     Minami-senjyu, Tokyo 116, Japan.}
  
\altaffiltext{18}{Department of Physics, Nagoya University, Nagoya 464, Japan.}
  
\altaffiltext{19}{National Astronomical Observatory, Mitaka, Tokyo 181, Japan.}
  
\setlength{\footnotesep}{\oldfootsep}
\renewcommand{\topfraction}{1.0}
\renewcommand{\bottomfraction}{1.0}     

\pagebreak


\vspace{-5mm}
\begin{abstract} 
\rightskip = 0.0in plus 1em
We present observations of microlensing event MACHO-98-BLG-35 which reached
a peak magnification factor of almost 80. These observations by the
Microlensing Planet Search (MPS) and the MOA Collaborations  place strong
constraints on the possible planetary system of the lens star and
show intriguing evidence for a low mass planet with a
mass fraction $4\times 10^{-5} \leq \epsilon \leq 2\times 10^{-4}$.  
A giant planet with $\epsilon = 10^{-3}$ is excluded from 95\% of 
the region between 0.4 and 2.5 $R_E$ from the lens star, where $R_E$ is
the Einstein ring radius of the lens.
This exclusion region is more extensive than the generic ``lensing zone"
which is $0.6 - 1.6 R_E$.  For smaller mass planets, we can
exclude 57\% of the ``lensing zone" for $\epsilon = 10^{-4}$  and  
14\% of the lensing zone for $\epsilon = 10^{-5}$. 
The mass fraction $\epsilon = 10^{-5}$
corresponds to an Earth mass planet for a lensing star of mass
$\sim 0.3 \msun$.  A number of similar events will provide statistically
significant constraints on the prevalence of Earth mass planets.   
In order to put our limits in more familiar terms, we have compared our
results to those expected for a Solar System clone averaging over possible
lens system distances and orientations. We find that such a system is
ruled out at the 90\% confidence level. A copy of the Solar System
with Jupiter replaced by a second Saturn mass planet can be ruled out at
70\% confidence.  Our low mass planetary signal (few Earth masses to 
Neptune mass) is significant at 
the $4.5\sigma$ confidence level.  
If this planetary interpretation is correct,  
the MACHO-98-BLG-35 lens system constitutes the first detection of
a low mass planet orbiting an ordinary star without gas giant planets.
~\footnote[1]{The data are available for interested readers in the following web 
sites. 

MPS~~~~~~~~~~{\bf~http://bustard.phys.nd.edu/MPS/98-BLG-35}

MOA~~~~~~~~{\bf~http://www.phys.vuw.ac.nz/dept/projects/moa/9835/9835.html}
}

\end{abstract}
\vspace{-5mm}
\keywords{gravitational lensing - Stars: low-mass, brown dwarfs}

\newpage
\section{Introduction}
\label{sec-intro}
Planetary systems in the foreground of the Galactic bulge or inner
Galactic disk form a class of gravitational
lens systems that can be detected through photometric microlensing 
measurements. These are multiple lens systems although in most cases,
the light curve is not easily distinguished from a single lens light curve.
A planet perturbs the gravitational potential of its
host star ever so slightly, and its effect may manifest itself as a brief
variation of the would-be single lens microlensing light 
curve \cite{mao-pac,gould-loeb,bol-falco}.
With sufficiently frequent and accurate observations, it is possible to 
detect planets with masses as small as that of the Earth \cite{benn-rhie96}
and measure the fractional mass $\epsilon$ and the (projected) distance 
of the planet $a$ from the host lens star.  This complements
other ground based extra-solar planet search techniques \cite{marcy-but-rev}
which have sensitivities that are not expected to extend much below the 
mass fraction of Saturn ($3\times 10^{-4}$). 

Planned space based observatories, such as the Space Interferometry Mission 
(SIM) \cite{sim} or the proposed Terrestrial Planet Finder (TPF) \cite{tpf},
Darwin \cite{darwin}, and Kepler \cite{kepler} satellites 
can not only detect planets as small as the
Earth, but they can also study a number of their properties. However, the
prevalence of low mass planets is not known, and the recent planetary 
discoveries via the radial-velocity technique \cite{marcy-but-rev}
suggest that our current understanding of planetary system formation
is incomplete. Circumstellar disks of less than a Jupiter mass around 
young stellar objects have been found with multiwavelength observations
\cite{padgett}. These could indicate the formation of planetary systems
without massive planets, but it is also possible that massive planets have
already formed in such systems.
Thus, microlensing can provide valuable statistical information
on the abundance of low mass planets that can be used to aid in the design
of these future space missions \cite{exnps-rep}.

The planetary signal in a gravitational microlensing event can be quite
spectacular due to the singular behavior of the caustics,
but the signal is always quite brief compared to the duration of
the stellar lensing event.   The small size of the caustic curves
($\approx \cal{O}(\sqrt{\epsilon})$) more or less determines the planetary
signal timescale.  The caustics of a planetary binary lens consist of the
stellar caustic (very near the stellar lens) and  planetary caustic(s), and
the detection probability  of the planet depends on the size and 
geometric distribution of the caustics which in turn depend on the 
fractional mass $\epsilon$ and the (projected) distance $a$ of the planet
from the stellar lens.   Theoretical estimations based upon a variety of 
``reasonable" detection criteria have shown that the 
detection probability of a planet with $\epsilon = 10^{-3}$ is
about 20\%  \cite{gould-loeb,wambsganss,gaudi-naber,di-scalzo,stefano-scalzo}, 
and for Earth-mass planets orbiting an M-dwarf primary
($\epsilon \approx 10^{-5}$), it is only about 2\% \cite{benn-rhie96}.    
Thus, microlensing planet search programs must generally observe
a large number of microlensing events in order to detect
planetary lensing events even if planets are ubiquitous.

For high magnification events with peak magnification $\Amax \gsim 20$,
the stellar caustic can cause planetary perturbations to the lightcurve, 
and the probability of detecting planets is very high \cite{griest-saf}.
For example, the detection probability for a giant planet in the 
``lensing zone" is close to $100\%$ for a microlensing event like 
MACHO-98-BLG-35 where the peak magnification was about 80.
High magnification occurs when the impact distance
is much smaller than the Einstein ring radius ($1 \gg \umin\simeq 1/\Amax$),
so the source comes very close to the location of the
stellar caustic. ($\umin$ is the impact distance in units of the
Einstein ring radius.)
If we recall that a single lens (stellar lens only) has a point caustic
at the position of the lens, the stellar caustic of a planetary binary
lens can be considered as this point caustic extended to a finite size
due to the gravitational perturbation of the planet. 
For a large range of planetary mass fractions
and separations, the stellar caustic will be perturbed by the planet,
and this will be visible near the lightcurve peak of a high magnification
event, whose timing can be predicted fairly accurately. 
Thus, high magnification events offer the opportunity to detect
a planet in a large range of locations in the vicinity of the lens star.
High magnification events are relatively rare, with a probability
$\sim \umin$, but when they occur, they should be observed
relentlessly.

Event MACHO-98-BLG-35 was the highest magnification microlensing event
observed to date, and it was one of the first high magnification events
that was closely monitored for evidence of planets near peak magnification
(see also Gaudi et al.~1998).
In this paper, we present a joint analysis of the
MACHO-98-BLG-35 data from the Microlensing Planet Search
(MPS) and MOA collaborations. This analysis
yields evidence consistent with a planet in the mass range
$4\times 10^{-5} \leq \epsilon \leq 2\times 10^{-4}$ which would be the
lowest mass planet detection to date, save the planetary system of
pulsar PSR B1257+12 \cite{pulsarplan}. We also compare our data to
binary lens
light curves for planetary mass fractions from $\epsilon = 3\times 10^{-7}$
to $\epsilon = 10^{-2}$ with separations ranging from $a = 0.2$ to
$a = 7.0$ measured in units of the Einstein ring radius of the total mass, 
$R_E$.  (From here on, $a$  is understood to be dimensionless, measured 
in units of $R_E$, unless stated otherwise.)
We find that giant planets are excluded over a large range of separations,
while there are also significant constraints extending down below an Earth
mass ($\epsilon \sim 10^{-5}$).

This paper is organized as follows.  Section
\ref{sec-alert} gives the chronology of the microlensing alerts for
MACHO-98-BLG-35 and a description
of the observations and data reduction. Section \ref{sec-srclen} discusses
the properties of the source and the lens.
Section \ref{sec-plan} describes our search for planetary signals 
in these data, and we discuss our conclusions in Section \ref{sec-con}.

\section{Alerts, Observations and Data Reduction}
\label{sec-alert}

Event MACHO 98-BLG-35 was discovered by the MACHO alert system~\footnote[1]{
Information regarding ongoing microlensing events can be obtained from the
EROS, MACHO, MPS, OGLE and PLANET groups via the world wide web:

EROS~~~~~~~~{\bf~http://www-dapnia.cea.fr/Spp/Experiences/EROS/alertes.html}

MACHO~~~{\bf~http://darkstar.astro.washington.edu/}

MPS~~~~~~~~~~{\bf~http://bustard.phys.nd.edu/MPS/}

OGLE~~~~~~~{\bf~http://www.astrouw.edu.pl/$\sim$ftp/ogle/ogle2/ews/ews.html}

PLANET~~{\bf~http://www.astro.rug.nl/$\sim$planet/index.html} }
\cite{macho-alert}
at a magnification of about 2.5 and announced at June 25.8, 1998 UT.
The MPS collaboration began observing this event with the
Monash Camera on the 1.9m telescope at Mt.~Stromlo Observatory (MSO) 
on the night of June 26, and had obtained 28 R-band
observations by July 3.6. Analysis of the MPS data set indicated
that this event would reach high magnification on July 4.5 with 
a best fit maximum magnification of $A\sim 33$. This was announced by
MPS via email and the world wide web.  This announcement called 
attention to the enhanced planet detection probability during high 
magnification.  The MOA collaboration responded to this alert and 
obtained a total of 162 observations with 300 second exposures
in the MOA custom red passband \cite{abe,reid}
from the  61-cm Boller and Chivens telescope at the Mt.~John 
University Observatory in the South Island of New Zealand over the 
next three nights. MPS obtained 35 more R-band observations over the next
three nights (the night of July 5 was lost due to poor weather at
Mt.~Stromlo) and then 65 additional observations of MACHO 98-BLG-35 over the
next two months as the star returned to its normal brightness. The MPS
exposures were usually 240 seconds, but they were reduced to 120 seconds 
near peak magnification to avoid saturation of the MACHO 98-BLG-35 images.

The MPS data were reduced within a few minutes of acquisition
using automated Perl scripts which call a version of the SoDOPHOT 
photometry routine \cite{sod}. During the night of July 4, near peak
magnification, the photometry of MACHO 98-BLG-35 was monitored by MPS
team members with a time lag of no more than 15 minutes after image
acquisition.  At approximately July 4.75 UT, a slight
brightening of the MPS measurements with respect to the expected single lens
light curve was noted. Shortly thereafter, MPS commenced more frequent
observations of MACHO 98-BLG-35, and it was followed as long as
possible, even at a very high airmass. Observations were obtained 
until July 4.801 UT at airmasses up to 3.64, but the high airmass data
are relatively noisy.

The MOA data were also reduced on-site using the fixed position version of
the DOPHOT program \cite{dophot} which is very closely related
to the SoDOPHOT routine used to reduce the MPS data. Both the MPS and MOA
photometry are normalized to a set of nearby constant stars using techniques
similar to that of \citeN{honeycutt}

The measurement uncertainties used in the analysis that follows are the 
formal uncertainties generated by the DOPHOT and SoDOPHOT photometry codes
with a 1\% error added in quadrature to account for flat-fielding and 
normalization uncertainties that are not included in the DOPHOT and SoDOPHOT
formal error estimates.

\section{Source and Lens Characteristics}
\label{sec-srclen}

Although this event is nominally a Galactic bulge event, it is actually located
at galactic coordinates of $l = 9.5435^\circ$, $b = -2.7757^\circ$ which is
toward the inner Galactic disk and outside the bulge. The unmagnified source
brightness reported by the MACHO Collaboration is $V = 20.7$, $R = 19.6$.
Our fit suggests that $\sim 10$\% of the light is due to an unlensed source
in the same seeing disk, so the magnitudes of the lensed source are closer to
$V = 20.8$ and $R = 19.7$. A crude estimate of the extinction can be
obtained from the dust map of \citeN{sfd-dust} (SFD)
which gives $E(B-V) = 1.46$, $A_V = 4.72$ and $A_R = 3.84$. However, SFD
were unable to remove IR point sources from their maps at such low Galactic
latitudes, so this probably overestimates the amount of extinction.
\citeN{stanek-dust} has investigated the SFD dust maps at low Galactic
latitudes and finds them to be highly correlated with the extinction
determined by other means. He suggests that the extinction at low latitudes
is roughly a factor of 1.35 lower than the SFD values. This gives
$E(B-V) = 1.08$, $A_V = 3.49$, and $A_R = 2.84$. If we use Stanek's method to
estimate the reddening, we get unreddened values of $V=17.3$ and $V-R=0.4$.
This is consistent with a G5 turnoff star of 2-3$\rsun$ near the Galactic
center or a solar type main sequence star at about $3\kpc$. However, 
the microlensing optical depth is quite small for a source star at $3\kpc$, 
so this is quite unlikely. Another possibility is that the source is a 
G5 horizontal branch
star at about the distance of the Sagittarius Dwarf Galaxy ($\sim 22\kpc$).

This estimate is quite sensitive to the assumed color of the star. If we
take $V-R=0.2$ as the dereddened color, then the source star is consistent
with
an early F main sequence star of 1.2-1.3$\rsun$ near or slightly beyond the
Galactic center. An error of 0.2 in $V-R$ is probably within the MACHO
calibration uncertainties for fields near the Galactic center \cite{calib}.
In any of these cases, the finite angular size of the 
source star is not likely to have a detectable effect on the shape of the
microlensing light curve.

The likely characteristics of the lens star depend somewhat on the location
of the source star. If the source and the lens are both located on the
near side of the Galactic center, then the source and lens share much of 
our galactic rotation velocity so that there is a small relative velocity 
between the lens and the line of sight to the source. This will generally
result in a long timescale event. On the other hand, if the source and lens
are on opposite sides of the Galactic center, then there will be a large
relative velocity between the lens and the line of sight to the source
resulting in a relatively short event. Because of this, the distribution of
event timescales in a low latitude Galactic disk field is rather broad, and
this makes it difficult to estimate the mass of the lens from the timescale
of the event. It is probably more accurate to estimate the mass of the lens
star from our knowledge of the mass function of stars in the Galaxy. This
would put the most likely mass at $\sim 0.3\msun$ with an uncertainty of
a factor of two or three. Because this event has a relatively short timescale,
it is likely that the lens and source are on opposite sides of the Galactic
center.

The observable features of microlensing events depend on the Einstein
ring radius $R_E$ which is given by
\begin{equation}
R_E^2 = {4GM\over c^2} {D_{o\ell}D_{\ell s}\over D_{o\ell}+D_{\ell s}}
\label{eq-re}
\end{equation}
where $D_{o\ell}$ and $D_{\ell s}$ are the observer-lens and lens-source
distances and $M$ is the total mass of the lens system. Eq. (\ref{eq-re})
gives the Einstein ring radius as measured at the distance of the lens,
so the angular size of the Einstein ring radius is $\alpha_E = R_E /D_{o\ell}$.
The Einstein ring radius is the characteristic length scale for gravitational 
microlensing, and $R_E$ is a few AU for typical Galactic microlensing events.
For event MACHO-98-BLG-35, we have determined the expected distribution of 
$R_E$ assuming a standard Galactic disk model of scale length $3\kpc$,
scale height $0.3\kpc$ with an assumed distance of $8\kpc$ to the Galactic
center. This gives $R_E = 3.2{+0.9\atop -1.1}\,$AU for a $1\msun$ lens
with a $2\sigma$ uncertainty extending from $1.1\,$AU to $5.0\,$AU. For a
more likely lens of $M=0.3\msun$, we have $R_E = 1.8{+0.5\atop -0.6}\,$AU 
with a $2\sigma$ uncertainty region of 0.6-$2.8\,$AU.  

The Einstein ring radius of the total mass of the lenses, $R_E$, is the
size of the ring image that occurs when the lens masses, the source star,
and the observer are aligned.  The angular position of an image is
nothing but the direction of the propagation vector of the photon beam
arriving at the observer from the source star. The angular position of 
the source is the position of the image of the source star when there is
no intervening lensing mass.   When the gravitational scattering angle is
small as in microlensing ($\sim 1$ mas), the angular positions in the sky
can be replaced by linear variables on a plane that is tangent to the
spherical surface of the sky.  Here, we have chosen the plane to be at 
the distance of the center of the lensing masses.  This plane is conventionally
referred to as the lens plane.   If we consider 
the lens plane as a complex plane, and we let $z$ and $\omega$ be the (complex)
position variables of an image and its source respectively,   
the binary lens equation is written as follows.
\begin{equation}
 \omega = z - R_E^2 \left({1-\epsilon \over \bar z - x_1}
                         + {\epsilon \over \bar z - x_2} \right) \ , 
\label{eqBilens}
\end{equation}
where the planetary lens of a fractional mass $\epsilon$ is at $x_2$ and the 
stellar lens is at $x_1$ on the lens axis chosen along the real axis of the
complex plane.   One can see that $R_E$ is a scale parameter of the lens plane,
and we  choose $R_E$  as the unit distance of the lens plane, which is a 
usual practice.   
\begin{equation}
   1  =  R_E 
\end{equation}
The lens equation (\ref{eqBilens}) shows that the lens parameter  space is given
by the fractional mass $\epsilon$ and the separation $|x_1 - x_2|$.  It is 
worthwhile to reflect that the source position variable $\omega$ is defined 
on the lens plane (at a distance $D_{o\ell}$ here) not on the plane that passes 
through the physical position of the source  at the  distance of $D_{o\ell}+D_{\ell s}$.   
If we call the lens plane parameterized by the source position variable $\omega$
the source plane and the lens plane parameterized by the image position variable
$z$ the image plane, the lens equation is an explicit mapping from the image plane
to the source plane; or, the lens equation is a mapping from the lens plane to
itself.

\section{Search for Planetary Signals}
\label{sec-plan}

The combined MPS and MOA data can be used to explore the possibility of
planetary companions to the lens star in two different ways. First,
we can fit the combined light curves with planetary lens models, and
compare the planetary lens light curves with the best fit single lens
light curve. As we show in subsection
\ref{subsec-sig}, there is a set of planetary lens models
that give a better fit to the data than the best single lens fit. However,
the apparent planetary signal is weak enough so that the planetary
parameters cannot be uniquely determined.
In addition to this possible planet detection, we can also rule out
a variety of possible planets orbiting the MACHO-98-BLG-35 lens
with sensitivity extending down to about an Earth mass.
This is discussed in subsection \ref{subsec-lim}, and we extend this
discussion to consider Solar System analogs in \ref{subsec-solsys}.

\subsection{Planetary Signal}
\label{subsec-sig}

We have fit the combined MPS and MOA light curves using the binary
lens fitting code described in \citeN{mps-98smc1}. We are able to
detect and characterize a significant deviation from a single lens
light curve near the peak magnification of this event.
The source's close approach to the angular position of the star and
the so-called stellar caustic results in both a very large
magnification and a substantial chance to detect a planetary
companion of the lens star \cite{griest-saf}. 
A planet will always extend the stellar caustic to a finite size, which 
changes the shape of the light curve at very high magnification and
accounts for the higher planet detection probability.
This is an advantage, but
it also has the consequence that the planetary lens parameters
are more difficult to determine for planetary deviations observed only at
high magnification.

The microlensing fit parameters that
pertain to both single and binary lenses are:
the Einstein radius crossing time, $t_E$, and
the time, $\t0$, and distance, $\umin$, of the closest approach between the 
line of sight to the source star and the lens system center of mass. $\umin$
is measured in units of the Einstein ring radius. In addition, there
are three parameters intrinsic to the binary lens fits: the binary lens
separation, $a$ (in Einstein ring radius units), the planetary lens mass
fraction, $\epsilon$, and the angle, $\theta$, between the source 
trajectory and the line connecting the lens positions. For $\theta=0$, the
source will approach the planet before it approaches the lens star.

\begin{figure}
\plotone{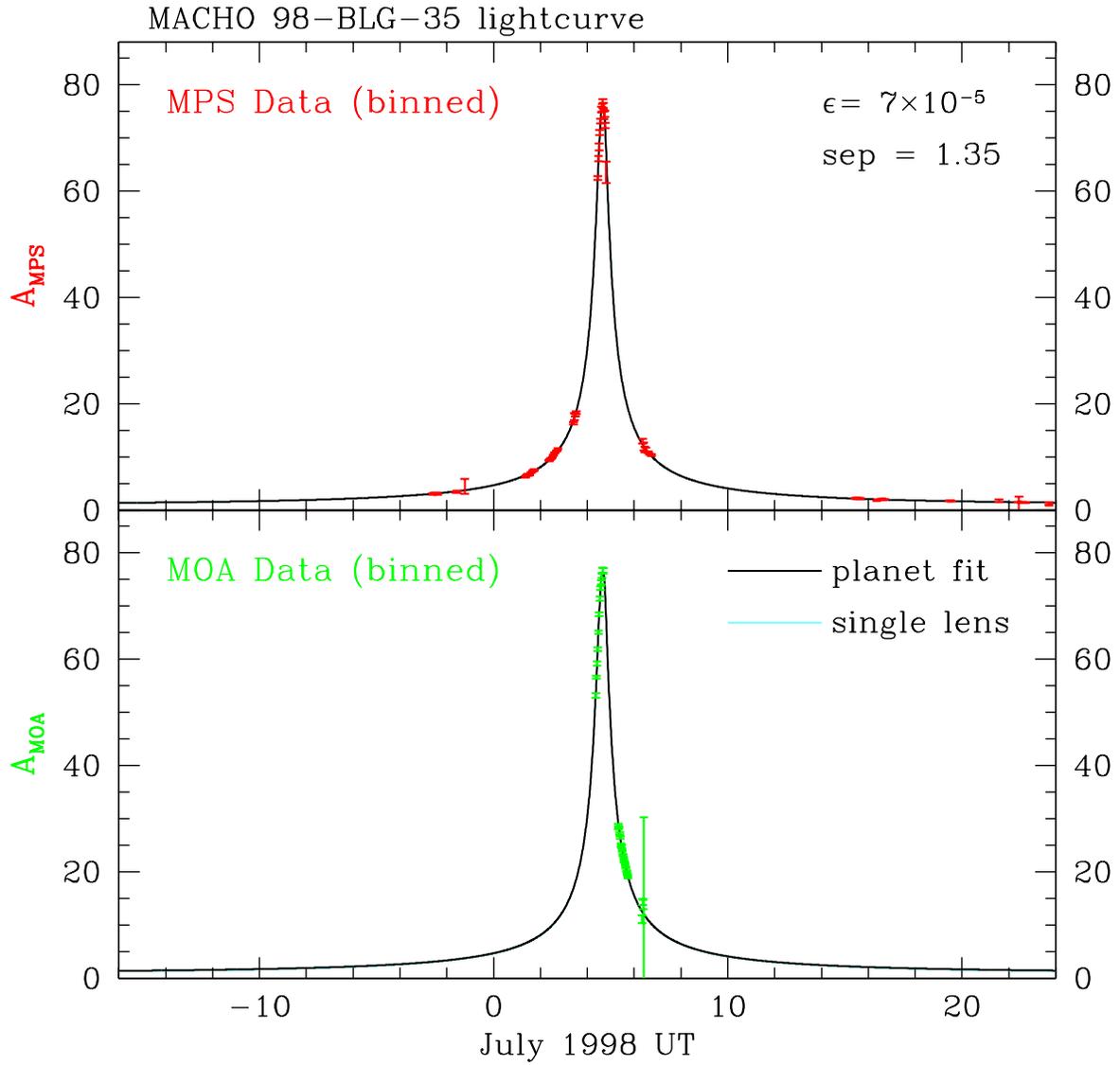}
\caption
{The light curves of the MPS and MOA data sets are plotted as
a function of time with the best fit planetary and single lens
light curves (which are nearly indistinguishable in this Figure. 
The data are binned and averaged on 0.03 day intervals, but the fits shown
are the best fits to the unbinned data.
  \label{fig-lcfull} }
\end{figure}

\begin{figure}
\plotone{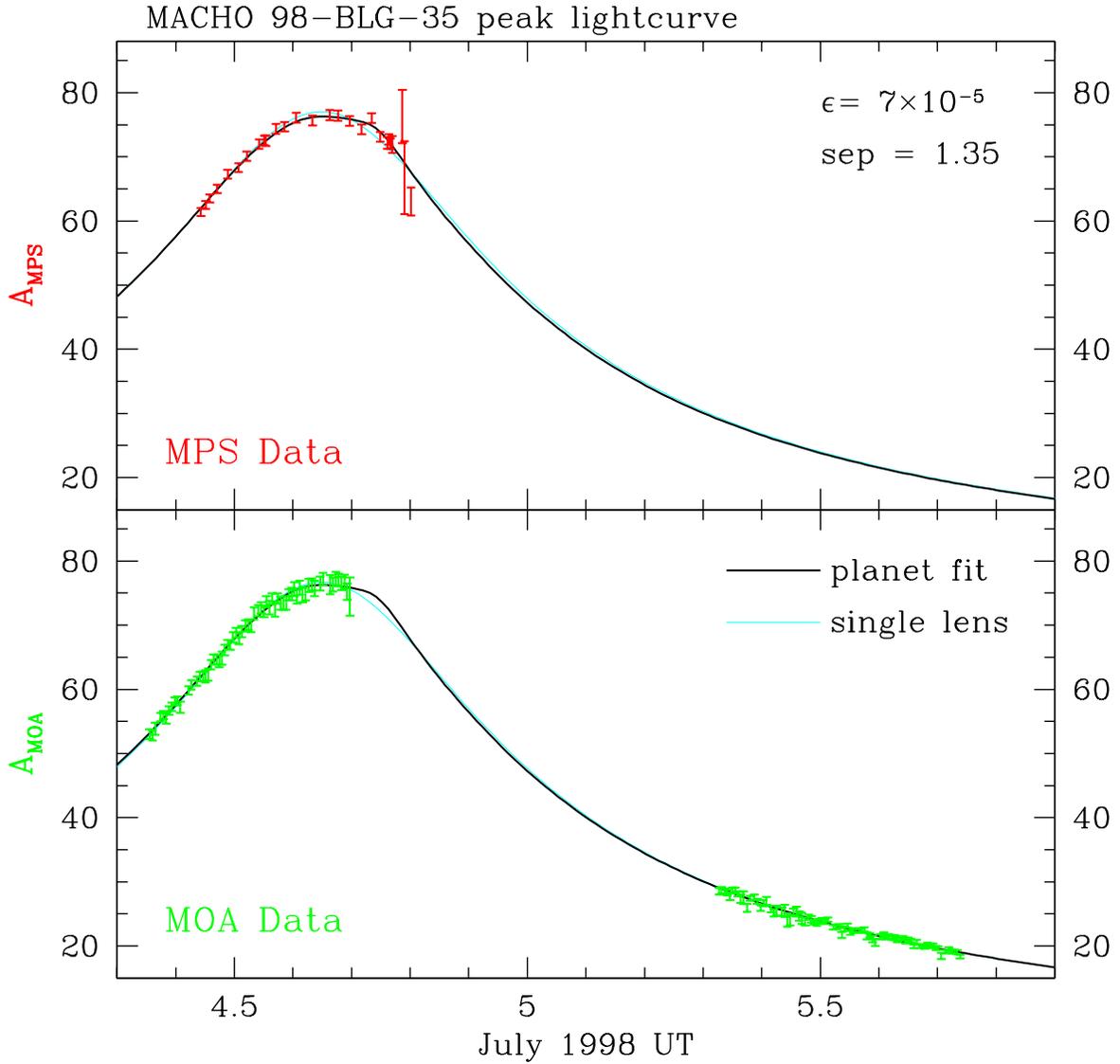}
\caption
{A close-up view of the MPS and MOA light curves near peak
magnification. The best fit planetary and single lens light curves
can be distinguished near peak magnification. The data are not binned.
  \label{fig-lc2d} }
\end{figure}

\begin{figure}
\plotone{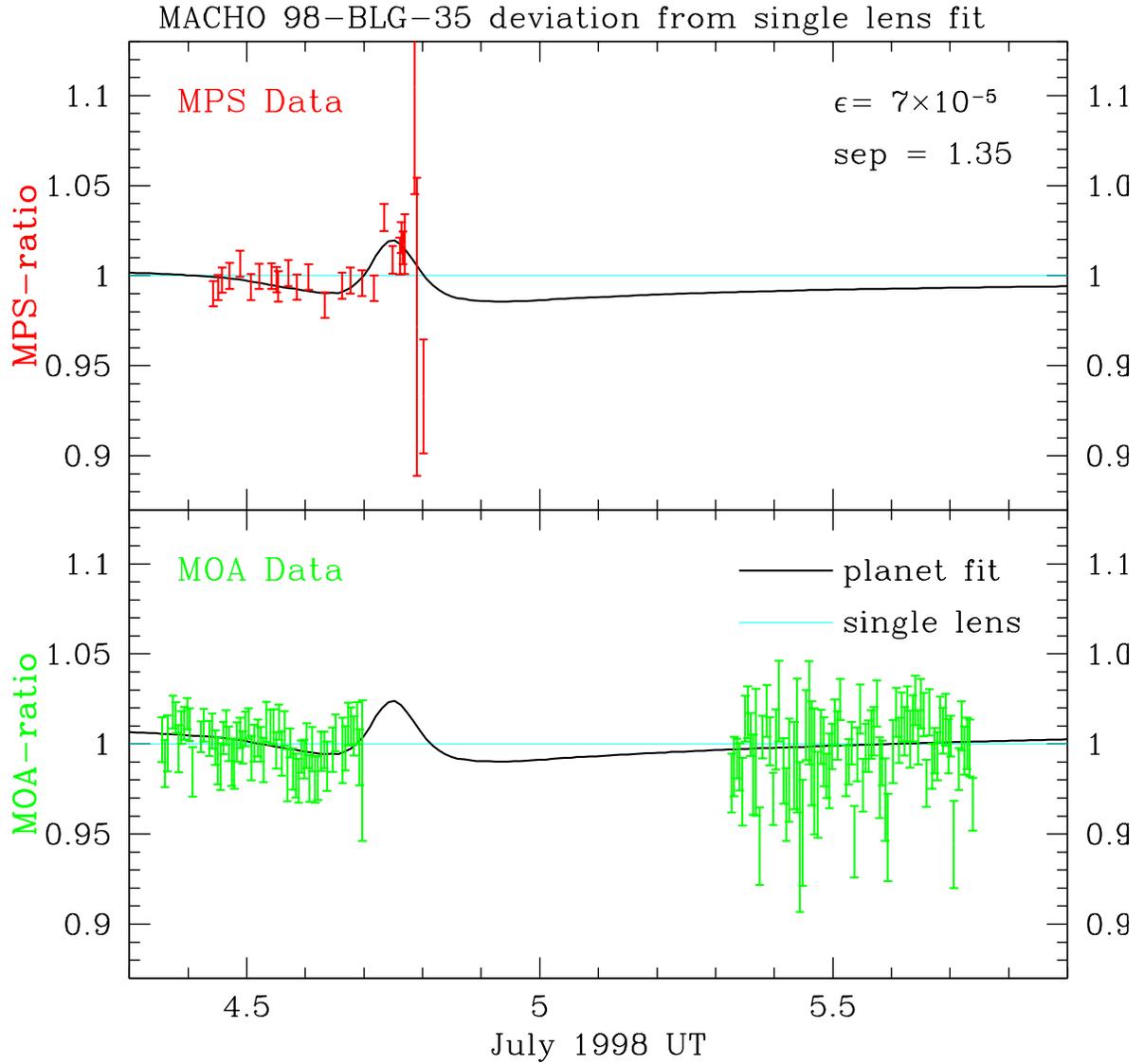}
\caption
{ The ratio of the data and the best fit planetary lensing light curve
to the best fit single lens light curve are plotted as a function of time.
The differences between the MPS and MOA fit light curves are due to the
different amounts of lensed and unlensed flux in the different fits.
  \label{fig-ratnb} }
\end{figure}

\begin{figure}
\plotone{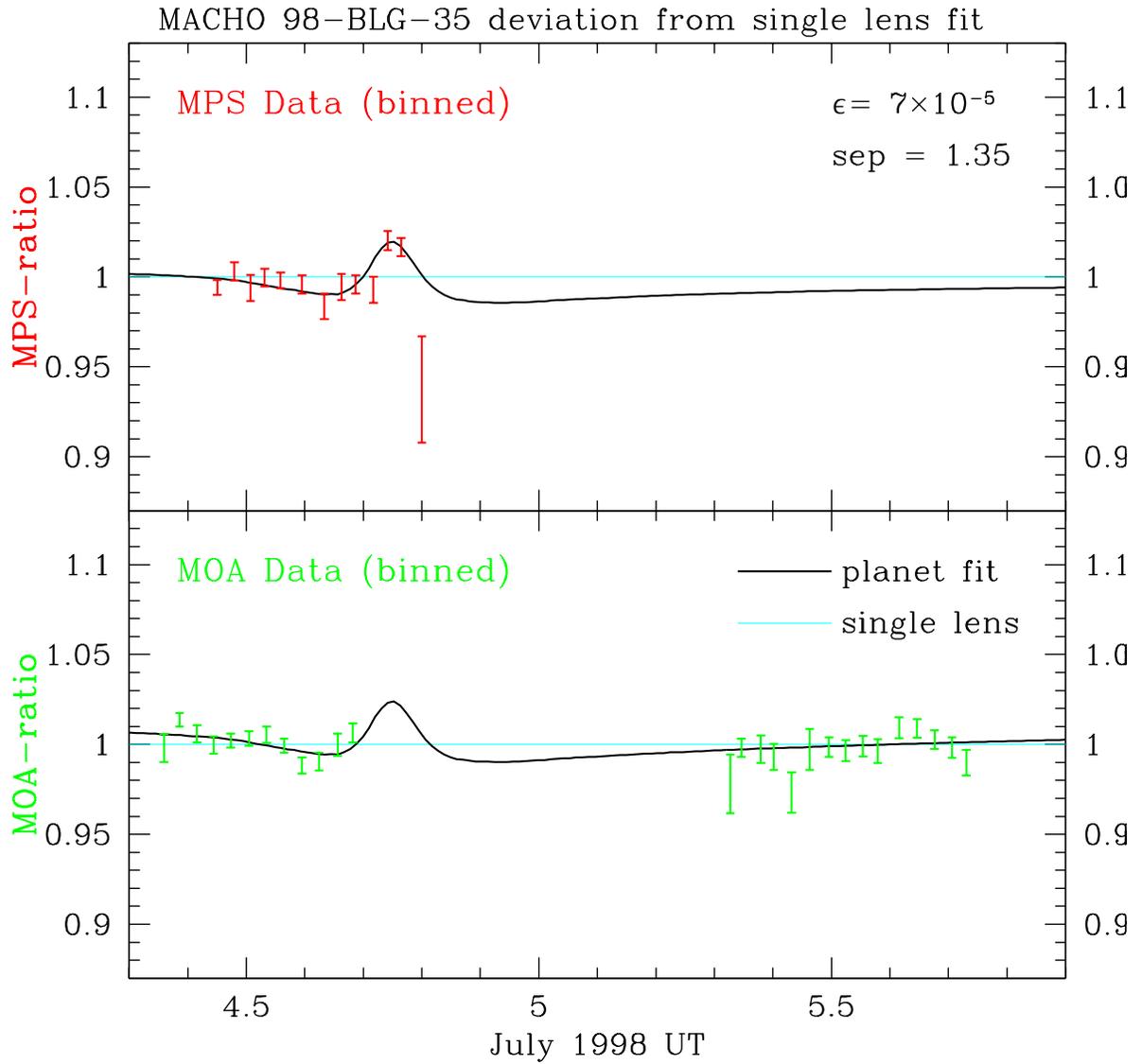}
\caption
{The same as Figure \ref{fig-ratnb}, but with
data binned and averaged on 0.03 day intervals. The the microlensing fits
have all been done with the unbinned data.
  \label{fig-ratb} }
\end{figure}

The parameters for three fit planetary microlensing
light curves and the best fit single
lens light curve are presented in Tables \ref{tab-lpar}-\ref{tab-bpar}. 
Figures \ref{fig-lcfull}-\ref{fig-ratb} show a comparison of the best
fit planetary lens light curve to the best fit single light curve. Figures
\ref{fig-lcfull} and \ref{fig-lc2d} show the data along with these
two light curves, while Figures \ref{fig-ratnb} and \ref{fig-ratb} show the
light curves and data divided by the best fit single lens light curve.
Because of the high frequency of observations, the data shown in Figures
\ref{fig-lcfull} and \ref{fig-ratb} have been averaged into 0.03-day long bins.
This binning is for display purposes only. All the fits have been done to the 
full data set.
The best fit light curve has a mass fraction of $\epsilon = 7\times 10^{-5}$
and has a $\chi^2$ improvement over the single lens fit of 23. This improvement
in $\chi^2$ (with three additional parameters)
implies that the planetary ``detection" is significant at the
$4.5\sigma$ level. This improvement in $\chi^2$ appears to be evenly divided
between the MPS and MOA data. 

The best fit $\chi^2 = 303.4$ for 275 degrees of freedom for the best fit
planetary lens curve. The probability
for a $\chi^2$ value at least this large is about 12\% assuming the
model is correct. For the best fit single lens curve, $\chi^2$ is larger by
23, but there are 3 more degrees of freedom because the model has fewer
parameters. The probability of a $\chi^2$ value this large to occur by
chance is only 2.4\%.

The fit $\chi^2$ is somewhat worse for the MPS data: $\chi^2 = 156.1$ for
118 degrees of freedom (assuming that 5 of the 10 fit parameters can be
associated with the MPS data). The probability that a $\chi^2$ value this
large will occur by chance is only about 1\%. However, there is one 
MPS observation from July 16 that contributes 19 to the $\chi^2$ value.
If this point is excluded, then the $\chi^2$ probability increases to 10\%.
This suggests that it is probably reasonable to use the SoDOPHOT and DOPHOT
generated error estimates.

Tables \ref{tab-lpar}-\ref{tab-bpar} also
present the fit parameters and $\chi^2$
values for both a ``low mass" and ``high mass" planetary fit in addition to
the best fit. These fits have $\chi^2$ values that are larger than the best
fit by about 4, so they correspond to approximate $2\sigma$ limits
on the planetary mass fraction. Thus, the $2\sigma$ constraint on 
$\epsilon$ is $1.7\times 10^{-5} < \epsilon < 7\times 10^{-4}$. The
$1\sigma$ limits are $4\times 10^{-5} < \epsilon < 2\times 10^{-4}$.
For a likely primary lens mass of 0.1-0.6$\msun$, the $1\sigma$ range
of planetary masses extends from about an Earth mass to twice the mass of
Neptune.  Table \ref{tab-fchi} indicates that the MPS
data prefer a lower planetary mass while the MOA data would prefer a somewhat
higher planetary mass.

Another apparent difference between the MPS and MOA data can be seen in
Table \ref{tab-bpar} which gives the best fit lensed ($F_\ell$) and 
unlensed ($F_u$) source flux values for each fit. These are given in
instrumental units, and it is necessary to include these parameters because
of the high stellar density in the fields where microlensing events are
observed. It is often the case that the stellar ``objects" detected by the
photometry codes will actually consist of several stars which are within
the same seeing disk. Only one of these stars will be lensed at a time,
so it will appear that only part of the flux of these stellar ``objects"
is lensed. In the case of MACHO-98-BLG-35, the MPS template frame was taken
when the source was magnified by about a factor of three, and the MPS
photometry code found three ``objects" with a brightness comparable to the 
unlensed brightness of the source within 2.5$\dpri$ of the 
source ``object\rlap."
These ``objects" were not separately resolved in the MOA template image 
which was taken when the source was magnified by a factor of 60. Presumably,
these stars were lost in the wings of the lensed source. This probably
accounts for the fact that the lensed flux ($F_\ell$) is about 92\% of the
total flux for the MPS data but only 66\% of the total flux for the MOA
data.  

\begin{figure}
\plotone{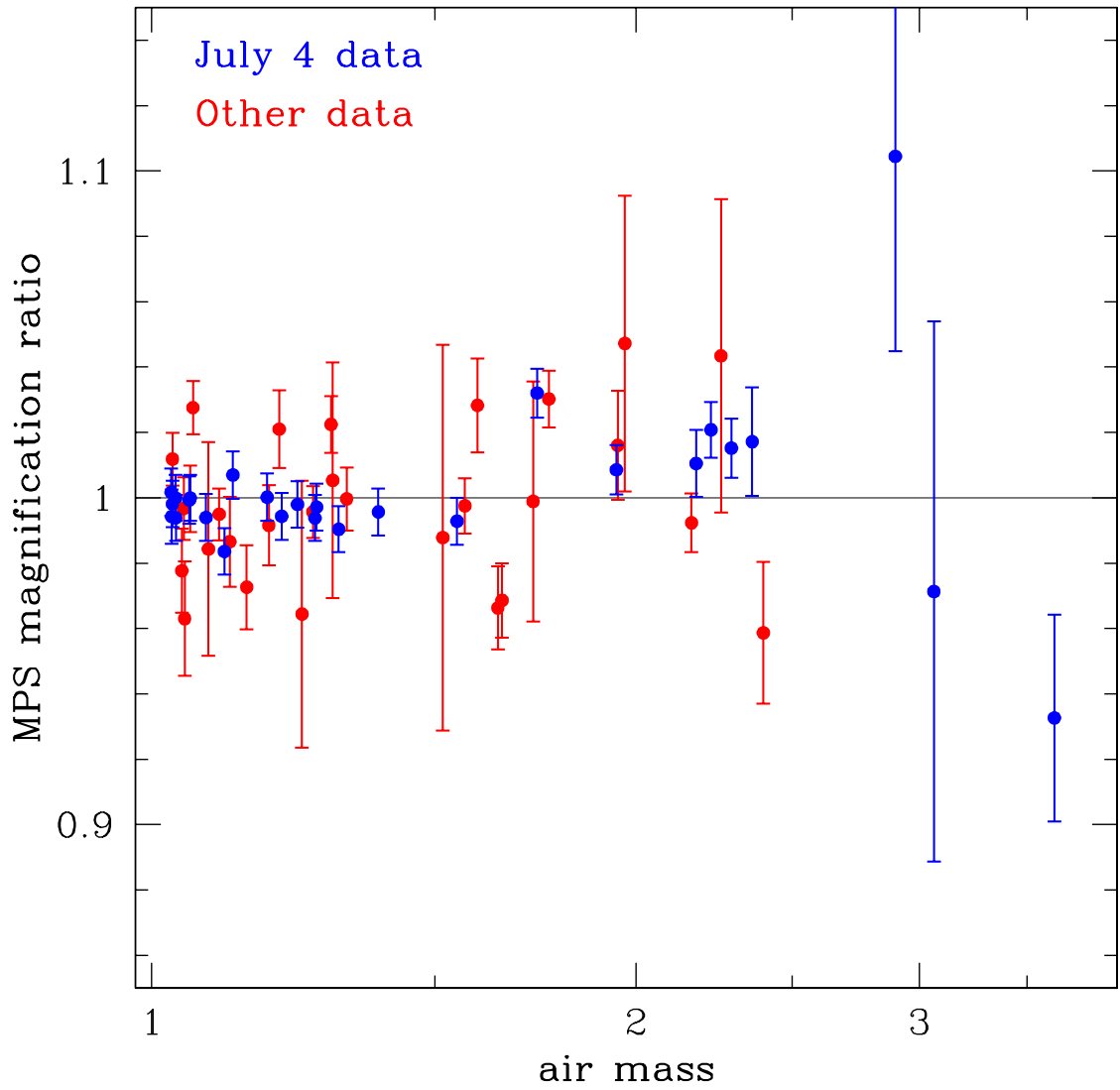}
\caption
{The magnification ratio from the best fit single lens light curve is
plotted as a function of airmass for the MPS data taken within the week
centered on the time of maximum magnification. The data from the night of
the planetary signal (July 4) are plotted in blue while the other data are
plotted in red.
  \label{fig-air} }
\end{figure}

The light curve features that are responsible for this apparent planetary 
detection can be seen most easily in Figures \ref{fig-ratnb} and 
\ref{fig-ratb}. The most significant deviations from the best single lens
fit are the 1.5\% decrease in flux relative to the single lens fit between 
July 4.34 and July 4.64 and then an increase of about 3\% to a relative
maximum at about July 4.75.    The slight flattening of the peak relative to 
the single lens fit seen in Figure{fig-lc2d} appears as the systematic trend
of declining from July 4.34 to July 4.64.  The increase of the flux seen 
in MPS data at about July 4.75 is  the sharpest feature seen
in the light curve, and it occurred while the source was setting from 
Mt.~Stromlo. The observations in the peak of this feature were taken at
an airmass ranging from 1.7 to 2.4 which is higher than most of our 
observations. (The final three observations had an airmass range of 2.8 to 3.6,
but the observations provide little weight to the planetary signal.)
Figure \ref{fig-air} shows the dependence of the magnification ratio on
airmass for the MPS data taken in the week centered on the peak magnification
for this event. There is a slight excess of observations at an airmass $\sim 2$
with a magnification ratio $> 1$, but as the color coding of the points
indicates, this is due to the fact that a large
fraction of these observations were taken during the light curve deviation
on July 4. Aside from this, there is no obvious trend with airmass, which
suggests that the feature seen in the MPS data is not a systematic error
due to the higher than average airmass of the observations.  A discussion of
the photometric accuracy of the MOA data can be found in an article by Yock
\cite{yock}. 

In addition to the planetary fits presented in Tables 
\ref{tab-lpar}-\ref{tab-bpar}, there is also a set of fits with parameters
very nearly identical to those in Table \ref{tab-lpar} except with the
planetary separation replaced by its inverse: $a\rightarrow 1/a$. This is
the well known ``duality" feature \cite{griest-saf} of the central caustic
for planetary lensing events, and it gives rise to a substantial uncertainty
in the separation of the planet from the lens. The best fit planetary
lens models have $a=1.35$ and $a=1/1.35$, but $a=1.45$ and $a=1/1.45$ are
consistent at $1\sigma$ while $a=2.07$ and $a=1/2.07$ are consistent
at a $2\sigma$ confidence level.

We should also consider the possibility that the light curve deviation is
caused by something other than a planet. For example, it is possible to get
a bump on the light curve from a binary source star lensed by a single lens
\cite{gaudi}.
Since the observed feature has a timescale about a factor of ten shorter than
the overall event timescale and an amplitude of about 3\% of the peak 
magnification, it might be possible to have a similar light curve if the
source star has a companion about 6 magnitudes fainter which has a peak
magnification 10 times larger than the factor of 80 observed for the primary
source star. This would give these observed features if the separation of 
two sources on the sky was about 0.012 Einstein radii. For typical lens
parameters and a random orientation of the source system orbit, this gives
a semi-major axis of 0.05 AU. 

However, if the source star is in a short period binary system, then the
trajectory of the source with respect to the lens system will not be
a straight line. The orbital motion of the source will generate a wobble
in the source trajectory which will cause periodic variations in the 
light curve \cite{xallarap}. No such variations are seen, and this
puts strong constraints on the nature of possible binary parameters of
the source star.  These variations may not be seen if the orbital period
of the binary source is larger than the timescale of the lensing event,
but this would require
that the source orbit be nearly edge on and that the two sources be just
passing each other at the time of peak magnification in order to reproduce such
a light curve. In addition, a secondary source 6 magnitudes fainter than 
the primary would probably have a very different color. Although MPS and MOA
have little color information for this event, other groups such as MACHO
and PLANET have observed it in different color bands.  In short, it would seem
to require several unlikely coincidences to have a binary source event mimic
a planetary perturbation in this case.
A future analysis including data from MACHO and
PLANET may be able to rule out this possibility.

\subsection{Planetary Limits}
\label{subsec-lim}

One of the benefits of these high magnification microlensing events is that
planets can be detected with high efficiency
at a large range of orbital separations around the
lens star. This means that the absence of a strong planetary signal can
be used to place limits on the possible planets of the lens system. We have
used the following procedure to quantify these limits. We consider a
dense sampling of the planetary lens parameter space with the planetary
separation, $a$, spanning
the range from $0.2$ to $7$ with an interval of $0.02$, $\theta$ varying from
0 to $2\pi$ at intervals of $1^\circ$, and $\epsilon$ ranging from
$3\times 10^{-7}$ to $10^{-2}$ in logarithmic intervals of $10^{1/8}=1.33$.
The other parameters were fixed so as to be quite close to the observed values
($\t0=4.65$ July UT, $t_E=20\,$days, and $\umin = 0.0125$).

For each set of parameters, an artificial light curve was created and imaginary
observations were performed at the same times and with the same error bars as
the actual MPS and MOA observations. The resulting artificial data set was
then fit with a 7-parameter single lens model, and the best fit $\chi^2$
value was determined. Since no photometric noise was added to these 
light curves, the fit $\chi^2$ values should be $< 1$ for events that
are indistinguishable from single lens events (at $1\sigma$ confidence)
or $> 1$ otherwise. The addition of Gaussian photometric noise should 
just add a contribution to $\chi^2$ equal to the number of degrees of 
freedom, so our measured $\chi^2$ values should be considered to be
the additional $\Delta\chi^2$ contribution caused by the planetary
signal. We set a detection threshold of $\Delta\chi^2 \geq 40$ which
corresponds to a $6.3\sigma$ deviation from the best fit single lens
light curve. Thus, we take
each set of planetary parameters that give best fit single lens curves
with $\Delta\chi^2 \geq 40$ to indicate that these planetary parameters have
been ruled out. The threshold of $\Delta\chi^2 \geq 40$ was selected to 
be somewhat larger than the deviation that we have actually detected in
the light curve.

All of these calculations were done using a point-source approximation to
calculate the planetary lensing light curves. This approximation is accurate
for most of the light curves, but some of the light curves will include caustic
crossings which would require a much more time consuming finite source
light curve calculation which would be complicated by the fact that we do not
know what the source size actually is. A reasonable estimate for the source
size projected to the plane of the lens system is 
$\lsim 0.004\,R_E$, so finite source effects are probably not very large. 
We have repeated our calculations for
finite sources with a much sparser sampling of the planetary lensing 
parameters. These calculations indicate that the point source calculations
slightly
underestimate the planetary detection probability for $\epsilon > 10^{-5}$,
but they overestimate the planetary detection probability for 
$\epsilon \lsim 10^{-5}$. Thus, our limits are conservative for 
$\epsilon > 10^{-5}$, but they may be overoptimistic for 
$\epsilon \lsim 10^{-5}$.

\begin{figure}
\plotone{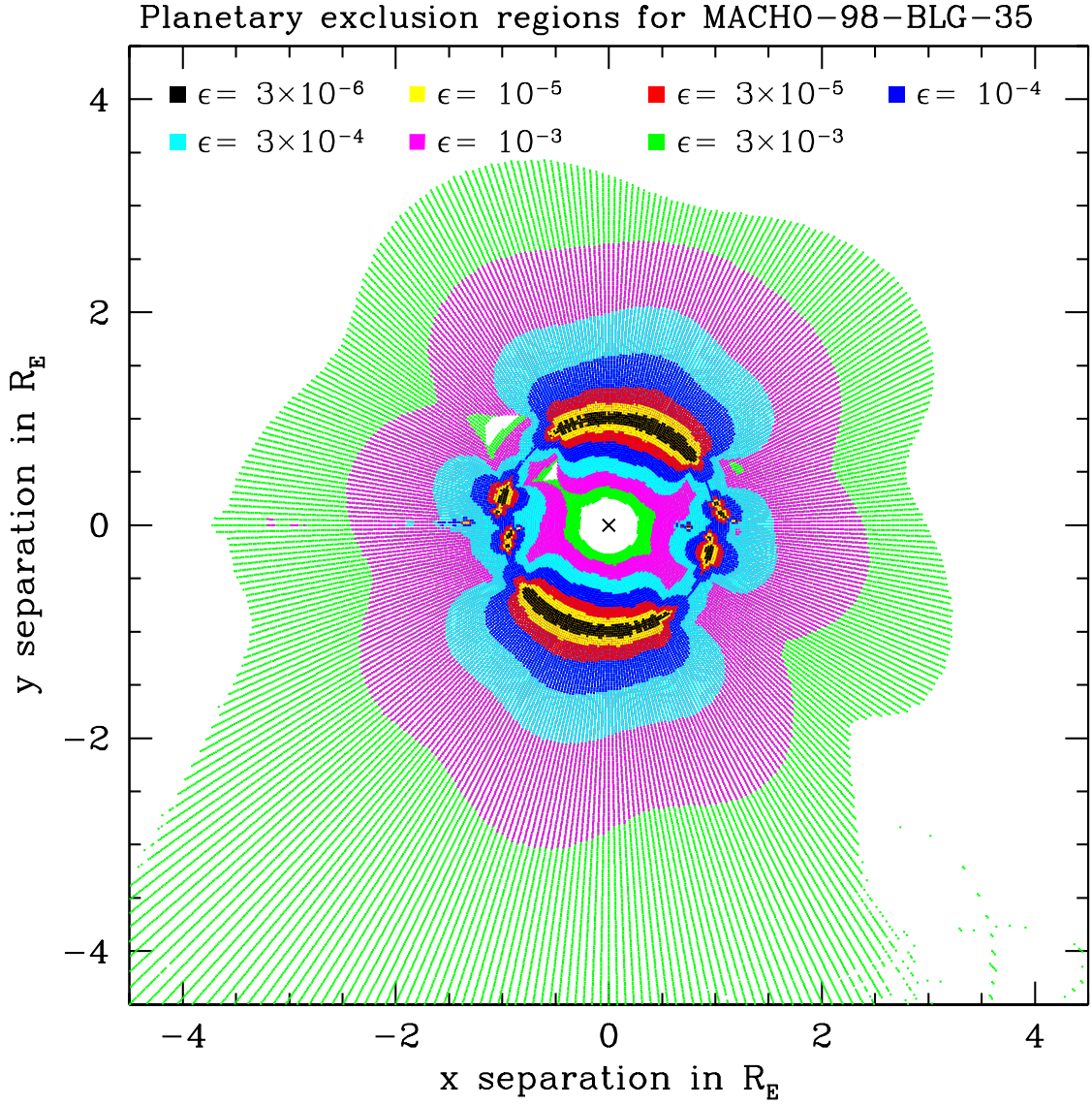}
\caption
{The excluded region of the lens plane is shown for a range of
planetary mass fractions ranging from $\epsilon = 3\times 10^{-6}$ to
$\epsilon = 3\times 10^{-3}$.
  \label{fig-excl2d} }
\end{figure}

Figure \ref{fig-excl2d} shows the regions of the lens plane where planets are
excluded for various planetary mass fractions ranging from 
$\epsilon = 3\times 10^{-6}$ to $\epsilon = 3\times 10^{-3}$. During the
lensing event, the source star crosses from right to left on the x-axis.
The gaps in the shaded regions represent our
lack of sensitivity to planets at particular $\theta$ angles where the
planetary light curve deviation occurs at a time when we have poor
coverage of the microlensing light curve.

Theoretical
papers on the microlensing planet search technique have generally asserted
that microlensing can detect planets that are in the so-called 
``lensing zone" which covers the range $0.6 < a  < 1.6$
\cite{gould-loeb,benn-rhie96,griest-saf}, but the 
exclusion regions for $3\times 10^{-4} \leq \epsilon \leq 3\times 10^{-3}$
in Figure \ref{fig-excl2d} clearly extend far beyond this region
\cite{dm96}.  There are
also significant exclusion regions for $\epsilon = 3\times 10^{-6}$ and
$\epsilon = 10^{-5}$ which correspond to planets of about an Earth mass,
so this event represents the first observational constraints on Earth-mass 
planets orbiting normal stars.

\begin{figure}
\plotone{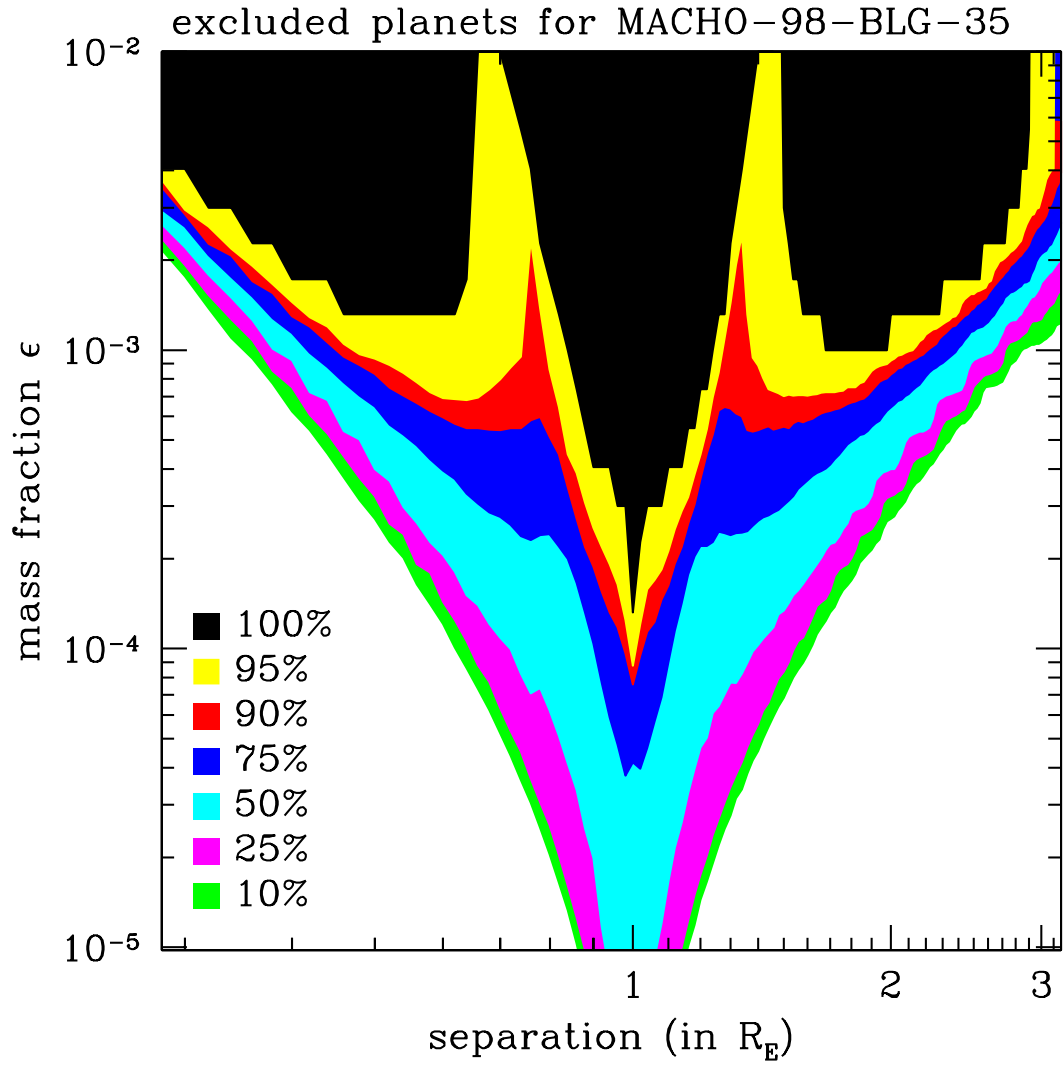}
\caption
{The parameters of planets excluded from the MACHO-98-BLG-35 system
are shown in the mass fraction - separation plane. The different colors
indicate what fraction of the planets with the given mass fraction and
separations are excluded.
  \label{fig-msex} }
\end{figure}

\begin{figure}
\plotone{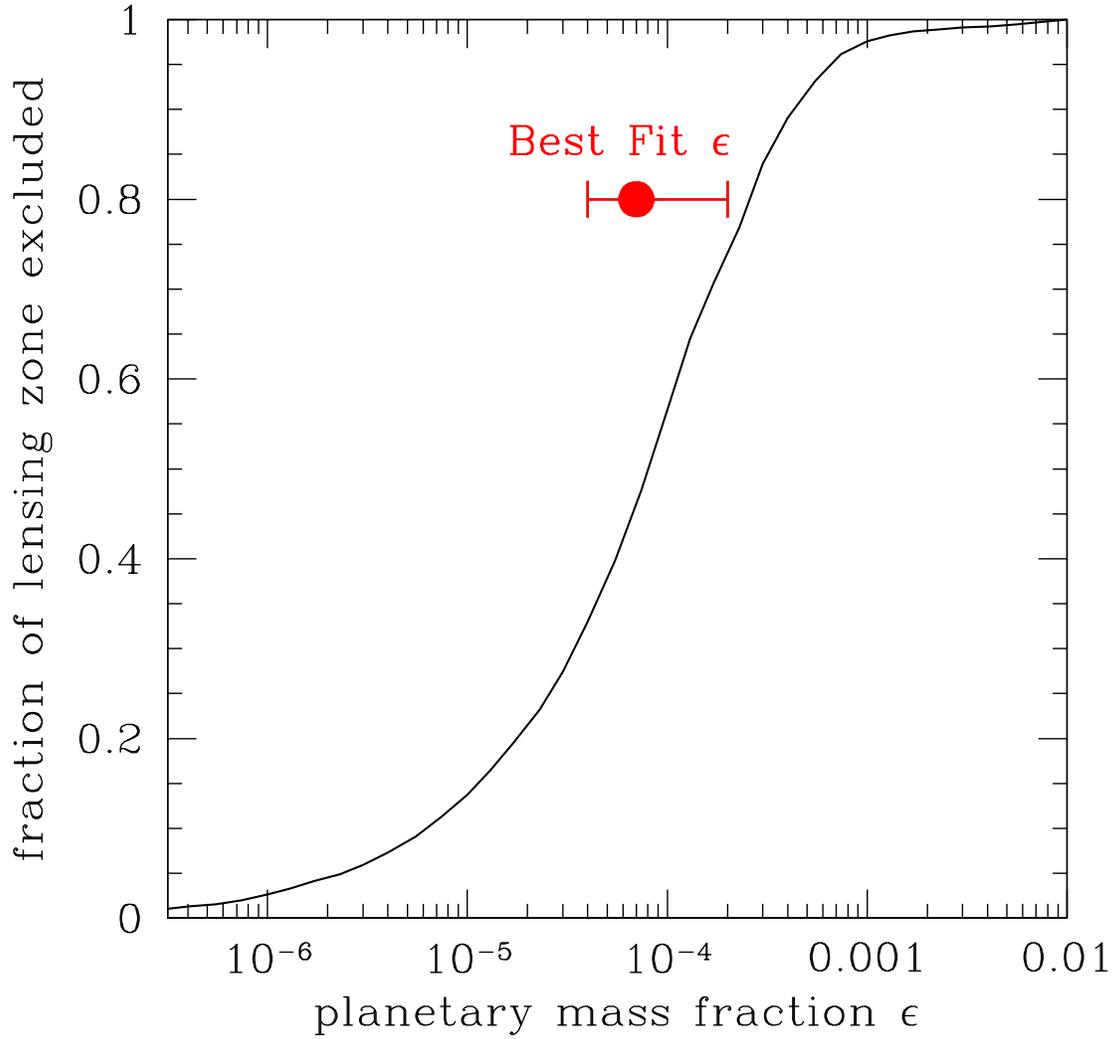}
\caption
{The fraction of the lensing zone, $0.6 \leq R_E \leq 1.6$, where
planets are excluded from the MACHO-98-BLG-35 system is plotted as
a function of the mass fraction $\epsilon$. The best fit mass fraction is
shown for comparison (plotted at an arbitrary $y$-axis value).
  \label{fig-zonex} }
\end{figure}

Another view of the planetary constraints can be seen in Figures
\ref{fig-msex}-\ref{fig-zonex}.
In Figure \ref{fig-msex}, we have averaged over all the $\theta$ values, and
we show the contours of the regions excluded at various confidence levels 
in the $a$-$\epsilon$ plane. The $x$-axis of Figure \ref{fig-msex} is
plotted on a logarithmic scale which reveals an approximate reflection
symmetry about $a=1$. This is an indication of the dual $a \rightarrow 1/a$
symmetry of light curves which approach the stellar caustic.
We make use of this duality property to construct Table \ref{tab-pexcl} 
which gives 50\% and 90\% confidence level exclusion ranges for the
planetary separation, $a$, as a function of the planetary mass fraction,
$\epsilon$. The limits of the planetary separation exclusion ranges are
chosen to be related by the $a \rightarrow 1/a$ transformation.
Table \ref{tab-pexcl} indicates that Jupiter-like planets
($\epsilon \geq 10^{-3}$) are excluded from a region much larger
than the usual lensing zone, while planets with 
$\epsilon \geq 3\times 10^{-5}$ (several Earth masses or more)
are excluded from a large fraction of the lensing zone.

Because of the enhanced planetary detection probability in the
lensing zone \cite{gould-loeb,benn-rhie96,griest-saf}, it is instructive 
to consider
the fraction of the lensing zone from which planets are excluded as a function
of mass fraction, $\epsilon$. This is plotted in Figure \ref{fig-zonex}, and
the $\epsilon$ range for our apparent planetary detection is shown as well.
This figure shows that the majority of the lensing zone must be free of
planets for $\epsilon \gsim 10^{-4}$, while more than 97\% of the
lensing zone can have no planets with $\epsilon \gsim 10^{-3}$. At
$\epsilon \sim 10^{-5}$, which corresponds to an Earth mass, more than
10\% of the lensing zone is excluded.

This relatively high planet detection probability is a feature of high
magnification events that was first emphasized by \citeN{griest-saf}.
The planet causes a distortion of the stellar caustic which can
be seen in the light curves of high magnification events for a large range
of planetary parameters, as we have shown. However, when only the stellar
caustic is detected, the determination of the planetary lens parameters
can be somewhat ambiguous if the planetary signal is not very strong.

Most of the detectable planetary microlensing 
signals are due to planetary caustics, and for these events, one can
determine $\epsilon$ and $a$ from the timing and
the magnification of the stellar peak with respect to the planetary 
deviation of the lightcurve \cite{gould-loeb,benn-rhie96,gaudi-gould}.
The planetary caustics  
cover a larger area of the lens plane than the stellar caustic does,
hence one expects a higher probability of a planetary discovery for a
planetary caustic event over a stellar caustic event.   However, 
the stellar caustic events have observational advantage that the
timing of the stellar caustic approach or crossing can  be predicted 
ahead of time which allows the
scheduling of additional observations that can greatly increase the planetary
detection probability.

\subsection{Solar System Analogs}
\label{subsec-solsys}

Thus far, we have discussed the limits placed on the planetary system of
the MACHO-98-BLG-35 lens star in terms of the units which are the most
convenient in the context of gravitational microlensing. We have talked about
the ``lensing zone" and used $R_E$ as our basic unit of distance. Since these
are the natural units of microlensing, this allows us to be precise and 
economical in our discussion of the limits, but they are not the units
that we usually use to measure solar systems. Although $R_E$ is typically
of order an AU, it does have a rather broad distribution. Thus, it might
be easier to see the significance of our limits if we translate them into
solar system units. To accomplish this, let us consider the possibility that
the lens system is a solar system analog. 
What are the chances that we would have
detected a light curve deviation with $\Delta\chi^2 > 40$ if the lens
star has a solar system just like that of the Sun?

In order to answer this question, we need to average over the parameters
of the lens system that are unknown. These include: the lens and source
distances, the inclination of the lens star's planetary plane, and the
position of each planet in its orbit. Also, since the lens star's mass is 
likely to be less than a solar mass, we need a prescription for how the
planetary separations scale with the mass of the lens star. In order to
simplify our calculations, we assume that the planetary separations scale
as $M^{1/2}$, but this does not have a large influence on our results.
(We continue to refer to planetary separations in AU, but it should be
understood that these distances are scaled as $M^{1/2}$. Thus, the
``Jupiter" of a planetary system orbiting a $0.3\msun$ star would be at
a orbital distance of 2.8 AU.)
We also assume that the distribution of fractional planetary masses
does not depend on the mass of the lens star. This means that our ``Jupiter"
will always have a mass fraction of $\epsilon = 10^{-3}$ independent
of the lens mass. We should also point out that our calculations are not
strictly correct for systems with more than one planet, since we have only
done calculations for binary lens systems. We have assumed that each planet
can be detected only if it could be detected in the same position in a
purely binary system. In practice, the additional lenses may increase the
light curve deviations for events near the detection threshold and push
these events above the detection threshold. So, our simplification
probably causes a slight underestimate of the planetary detection
probability.

Applying this procedure to our MACHO-98-BLG-35 data, we find that
a Solar System analog is excluded at the 90\% confidence
level. In 88\% of the cases,
the Jupiter-like planet ($\epsilon = 10^{-3}$ at 5.2 AU) would be 
detected, and the Saturn-like planet ($\epsilon = 3\times 10^{-4}$ at
9.5 AU) would be detected 19\% of the time although the Jupiter
would be also seen in most of these cases. About 1\% of the time
the Earth, Uranus or Neptune analogs would be seen.

If we modify the Solar System analog to replace the Jupiter-like planet
with a Saturn mass planet ($\epsilon = 3\times 10^{-4}$) at 5.2 AU,
we find that this system can be excluded at 69\% confidence. The Saturn
in Jupiter's orbit is seen 64\% of the time while both Saturns
can be detected in about 15\% of the cases.

Finally, let us consider planetary systems in which Jupiter and Saturn have 
been replaced by Neptune-like planets with $\epsilon=5\times 10^{-5}$.
This is a planetary configuration suggested by \citeN{peale} based on
consideration of planet formation theory, and following Peale, we also
introduce an additional Earth at 2.5 AU because the formation of such
a planet would be expected if Jupiter were absent. We find that there is
a 36\% chance that such a low mass planetary system would give rise to
an unacceptably large signal and be excluded by our data. About 29\% of the
time, the Neptune at Jupiter's position would be seen while the Neptune
in Saturn's orbit would be seen 4.5\% of the time and the Earth at 2.5 AU
would be seen 2.5\% of the time. As with the solar system analog, the
remaining planets contribute about 1\% of the total detection probability.
While the exclusion of such planetary systems at 36\% confidence is not
a very significant result by itself, it does indicate that with a few more
similar data sets, we will begin to be able to address the question of the
abundance of planetary systems without gas giants.
Of course, the apparent planetary signal in our data is consistent with the
detection of just such a system, so it is quite possible that we have
made the first detection of a planet in a system with no gas giants.

\section{Discussion and Conclusion}
\label{sec-con}

Our 90\% confidence level exclusion of a Solar System analog planetary system
is apparently the tightest constraint on planetary systems like our
own to date. The radial-velocity program of Marcy and Butler has
one or two systems in which they can constrain $\epsilon \sin i < 10^{-3}$
\cite{marcy-priv} at present. This is 
comparable to our constraint except for the
additional uncertainty due to the unknown inclination angle, $i$. However, 
their current radial-velocity sensitivity is good enough  to
obtain similar limits for hundreds of stars once they have high sensitivity
observations spanning the decade-long orbital periods of planets at 5 AU. 
So, within a few years, the radial-velocity groups will likely have 
similar constraints for hundreds of stars.

The real strength of the microlensing technique lies not with the ability to
detect Jupiter analogs, but in the sensitivity to lower mass planets.
Of course, low mass planets are more difficult to 
see with any technique, but with microlensing the planetary signals do not
get substantially weaker as the planetary mass drops as they do for other
techniques. Instead, the microlensing signatures of extra-solar planets become
{\it rarer} and {\it briefer} as the planetary mass decreases down to 
$\epsilon \sim 10^{-5}$ where finite source effects become important.
If we had 10 microlensing events
with limits similar to MACHO-98-BLG-35, then our statistical information on
the prevalence of Jupiter-like planets would be interesting, but probably
not competitive with what will be learned from the radial-velocity searches.
However, we would expect to detect several Neptune-mass planets with
a signal substantially stronger than the planetary signal seen in
our MACHO-98-BLG-35 data if most planetary systems have Neptune-mass planets.
Our sensitivity to Neptune and probably Saturn-mass planets would very
likely be beyond the sensitivity of the radial-velocity searches or 
other ground based planet search techniques. So, if it were the case that
most planetary systems have no planets more massive than Neptune, 
gravitational microlensing would likely be the only ground based planet
search technique sensitive to these planetary systems.

Our planetary results for event MACHO-98-BLG-35 have made use of the
large planet detection efficiency for high magnification microlensing
events that was first emphasized and quantified by \citeN{griest-saf}.
Although there are probably more detectable planetary signals for
low magnification events, the high magnification events have the advantage
that the planetary signal is expected near the time of peak magnification
which can be predicted with relatively sparse observations (once, or
preferably, a few times per day). If the light curve is then sampled
very frequently near peak magnification, there will be a high probability
of detecting a planet. The light curve can be sampled more sparsely
after the magnification decreases. Because we know when the planetary
signal is likely to occur, it is possible to discover a low mass planet with
a relatively small total number of observations.
It is critical, however, that the high magnification events be discovered
in advance and that they be observed frequently enough to predict their
peak magnification.

Let us consider the recent history of microlensing events discovered in
real time. The MACHO alert system has found 4 microlensing events with
a peak magnification $> 20$, in addition to a number of high magnification
binary lensing events with large mass fractions ($\epsilon > 0.1$).
The high magnification
events are MACHO-95-BLG-11, MACHO-95-BLG-30, MACHO-98-BLG-7, and
MACHO-98-BLG-35. OGLE has also found 4 such events:
OGLE-98-BUL-29, OGLE-98-BUL-32, OGLE-98-BUL-36
\footnote[2]{The high magnification was seen only in the MACHO data for 
OGLE-98-BUL-36, but this event was missed by the MACHO Alert 
system because MACHO
had data in only one color.}, and OGLE-99-BUL-5. Of these events, only
MACHO-95-BLG-30 \cite{macho95b30}, MACHO-98-BLG-35, and OGLE-99-BUL-5
were discovered substantially before peak magnification, while MACHO-95-BLG-11
and OGLE-98-BUL-29 were discovered in the 24-hour period preceding
peak magnification. Any steps that might be taken towards earlier discovery
of microlensing events in progress are likely to improve the planet detection
efficiency significantly. It is also critical that detected events be observed
frequently enough to predict high magnification events reliably.

The high efficiency for planet detection
for this event and the uncertainty in the planetary lensing parameters 
are both consequences of the very high magnification of lensing event
MACHO-98-BLG-35. 
More accurate planetary parameters can be obtained for
events where the planetary caustic is crossed or approached, which
generally occurs at a more modest magnification. Detectable
light curve deviations caused by an approach to the planetary caustic
are more frequent than light curve deviations caused by an approach to 
the stellar caustic, but we cannot predict in advance when a
planetary caustic might be approached so it requires more telescope
time to find such events.  
Dedicated microlensing follow-up programs
such as those being run by PLANET \cite{planet95} and MPS, are 
required in order to have a reasonable prospect of detecting such events. 
However, despite the complete longitude
coverage of the PLANET collaboration and the 1.9m telescope used by MPS,
the current generation of microlensing planet search programs is not
able to follow enough microlensing events with sufficient photometric
accuracy to obtain statistically significant constraints on the
abundance of low mass planets. This would require a more ambitious
microlensing follow-up program along the lines of that presented by
\citeN{peale}.

In summary, we have presented the first observational constraints on
a planetary system from gravitational microlensing, including a $4.5\sigma$
detection of an apparent planetary signal. The mass fraction of this
planetary companion to the lens star is likely to be in the range
$4\times 10^{-5} \leq \epsilon \leq 2\times 10^{-4}$.
Depending on the lens star mass, these mass fractions correspond to
planetary masses in the range from a few Earth masses up to about two
Neptune masses. Our data also place strong constraints on the planetary system
which may orbit the lens star. A system just like our own is excluded at 90\%
confidence while a system like ours with Jupiter replaced by a Saturn-mass
planet can be excluded at 70\% confidence. For a planetary system like
our own with Jupiter and Saturn replaced by Neptunes we would expect
a signal at least twice as strong as the one that we have detected about
30\% of the time. Our results demonstrate the sensitivity of the gravitational
microlensing technique to low mass planets. If we take our low mass planet
detection at face value, then it suggests that the most common planetary 
systems in the Galaxy may be the ones with their most massive planets less 
massive than a gas giant.

\acknowledgements
\section*{Acknowledgments}

We would like to thank the MACHO Collaboration for their early announcement
that made our observations of MACHO-98-BLG-35 possible, and we would also
like to thank the OGLE and EROS collaborations for their alerting us
to the ongoing microlensing events that they discover.

This research has been supported in part by the NASA Origins 
program (NAG5-4573), the National Science Foundation (AST96-19575), and
by a Research Innovation Award from the Research Corporation.  Work
performed at MSSSO is supported by the Bilateral Science and
Technology Program of the Australian Department of Industry,
Technology and Regional Development.  
Work performed at the University of Washington
is supported in part by the Office of Science and Technology
Centers of NSF under cooperative agreement AST-8809616.
The authors from the MOA group thank the University of Canterbury 
for telescope time, the Marsden and Science Lottery Funds of NZ, the 
Ministry of Education, Science, Sports and Culture of Japan, and the 
Research Committee of Auckland University for financial support, and
Michael Harre for discussions.

\clearpage

\begin{deluxetable}{lcccc}
\tablecaption {Binary lensing parameters and statistics \label{tab-lpar} }
\tablewidth{0pt}
\tablehead{
        \colhead {parameter} &
        \colhead {single lens}  &
        \colhead {best fit}  &
        \colhead {low mass}  &
        \colhead {high mass}
}
\startdata
    $\t0$ (July UT)
  & 4.65(9)
  & 4.65(9)
  & 4.65(9)
  & 4.66(9)      \nl
    $t_E$ (days)
  & 21.13(56)
  & 21.45(22)
  & 21.49(21)
  & 21.25(27)  \nl
   $\umin$
  & 0.01322(38)
  & 0.01299(14)
  & 0.01296(13)
  & 0.01208(16)  \nl
    $a$
  & 0
  & 1.35(3)
  & 1.19(2)
  & 2.07(8)   \nl
   $\theta$ (rad)
  & 0
  & 1.94(4)
  & 1.91(3)
  & 2.17(3)  \nl
   $\epsilon$
  & 0
  & $7.0(1.5) \times 10^{-5}$
  & $1.7      \times 10^{-5}$
  & $7.0      \times 10^{-4}$     \nl
   $\chi^2 / $ (d.o.f)
  & 326.45/278
  & 303.44/275
  & 307.04/275
  & 307.65/275  \nl
\enddata
\tablenotetext{} { The parameters of the best single lens fit are compared
with those of the best planetary binary lens fit along with ``low mass" and
``high mass" planetary fits which represent approximate $2\sigma$ limits on
the mass fraction.
  }
\end{deluxetable}
 
\begin{deluxetable}{lcccc}
\tablecaption {Fit $\chi^2$ values for individual data sets \label{tab-fchi} }
\tablewidth{0pt}
\tablehead{
        \colhead {data set} &
        \colhead {single lens}  &
        \colhead {best fit}  &
        \colhead {low mass} &
        \colhead {high mass}
}
\startdata
  MPS~R
  & 170.40/123
  & 157.25/123
  & 156.13/123
  & 163.16/123  \nl
  MOA~red 
  & 156.06/162
  & 146.19/162
  & 150.91/162
  & 144.49/162  \nl
\enddata
\tablenotetext{} { The $\chi^2$ values for the MPS and MOA data sets are
shown for the best single lens and planetary fits along with the $2\sigma$
upper and lower mass fraction fits. }
\end{deluxetable}
 
\begin{deluxetable}{lcccc}
\tablecaption {Lensed and unlensed fluxes ($F_\ell$ \& $F_u$)\label{tab-bpar} }
\tablewidth{0pt}
\tablehead{ 
        \colhead {data set} &  
        \colhead {single lens}  & 
        \colhead {best fit}  & 
        \colhead {low mass}  &
        \colhead {high mass}
}
\startdata
  MPS~R $F_\ell$
  & 199.2(4)
  & 195.4(4)
  & 194.9(4)
  & 197.0(4)       \nl
  MPS~R $F_u$
  & 13.6(1.4)
  & 16.5(1.4)
  & 17.2(1.4)
  & 16.3(1.4)      \nl 
  MOA~red $F_\ell$
  & 287.1(7)
  & 282.5(7)
  & 281.0(7)
  & 283.6(7)      \nl
  MOA~red $F_u$
  & 115(23)
  & 147(23)
  & 155(23)
  & 199(23)         \nl
\enddata
\tablenotetext{} { These fluxes are given in instrumental units which are
arbitrary, but are useful for comparing the different fits. }
\end{deluxetable}

\begin{deluxetable}{lcc}
\tablecaption {Planetary exclusion regions \label{tab-pexcl} }
\tablewidth{0pt}
\tablehead{
        \colhead {mass fraction: $\epsilon$} &
        \colhead {90\% excluded}  &
        \colhead {50\% excluded}
}
\startdata
  $3\times 10^{-3}$
  & $0.27 \leq a \leq 3.70$
  & $0.16 \leq a \leq 6.25$ \nl
  $ 10^{-3}$
  & $0.37 \leq a \leq 2.70$
  & $0.22 \leq a \leq 4.55$ \nl
  $3\times 10^{-4}$
  & $0.74 \leq a \leq 1.35$
  & $0.39 \leq a \leq 2.56$ \nl
  $ 10^{-4}$
  & $0.94 \leq a \leq 1.06$
  & $0.58 \leq a \leq 1.72$ \nl
  $3\times 10^{-5}$
  & $-$
  & $0.76 \leq a \leq 1.31$ \nl
  $ 10^{-5}$
  & $-$
  & $0.88 \leq a \leq 1.13$ \nl
  $3\times 10^{-6}$
  & $-$
  & $0.98 \leq a \leq 1.02$ \nl
\enddata
\end{deluxetable}

\clearpage

\end{document}